\documentclass[a4paper,12pt]{article}
\title {On the von Neumann equation with time-dependent Hamiltonian. Part II: Applications}

\author{Maciej Kuna$^\dagger$ and Jan Naudts$^\circ$ \\
\strut\\
\small          $^\dagger$Wydzia\l\ Fizyki Technicznej i Matematyki Stosowanej, Politechnika Gda\'{n}ska,\\
\small	  ul. Narutowicza 11/12, 80-952 Gda\'{n}sk, Poland\\
\small	  ~~E-mail: maciek@mifgate.mif.pg.gda.pl.\\
\small          $^\circ$Departement Fysica, Universiteit Antwerpen,\\
\small          Groenenborgerlaan 171, 2020 Antwerpen, Belgium\\
\small	  ~~E-mail: jan.naudts@ua.ac.be.
}

\date {}

\usepackage{amsfonts}
\usepackage{amsmath}
\usepackage[margin=1.0 in]{geometry}

\newcommand{\be}{\begin{eqnarray}}
\newcommand{\ee}{\end{eqnarray}}

\def\Io{{\mathbb I}}

\def\sn{\,{\rm sn}}
\def\cn{\,{\rm cn}}
\def\dn{\,{\rm dn}}
\def\sech{\,{\rm sech}}

{}
\newtheorem{proposition}{Proposition}{}
{}
{}

\begin{document}
\maketitle

\begin{abstract}
This second part deals with applications of a general method
to describe the quantum time evolution determined by a
Schr\"odinger equation with time-dependent Hamiltonian.
A new aspect of our approach is that
we find all solutions starting from one special solution.
The two main applications are reviewed,
namely the Bloch equations and the harmonic oscillator
with time-dependent frequency. Even in these well-known
examples some new results are obtained.
\end{abstract}

%%%%%%%%%%%%%%%%%%%%%%%%%%%%%%%%%%%%%%%%%%%%%%%%%%%%%%%%%%%%%%%%%%%%%%%%%%%%%%%%%
\section{Introduction}

The study of quantum models with a time-dependent Hamiltonian is relevant for many
branches of physics. Instead of solving the Schr\"odinger equation for a quantum
system together with its environment, it is often possible to replace the
action of the environment by a time-dependent term in the Hamiltonian of the system.
A slightly more general problem is that of solving the von Neumann equation
\be
\frac {{\rm d}\,}{{\rm d}t}\rho(t)=
i[\rho(t),H(t)],
\label {intro:vNeq}
\ee
which describes the time evolution of the density matrix $\rho$.
Its formal solution is
\be
\rho(t)=U(t)\rho(0)U(t)^*.
\ee
The unitary operators $U(t)$ satisfy the equation of motion
\be
i\left(\frac {{\rm d}\,}{{\rm d}t}U(t)\right)U^*(t)=H(t).
\label {intro:ueq}
\ee

A considerable body of knowledge exists about solving
these equations in the case that both the Hamiltonian and the density
matrix (minus an operator commuting with the Hamiltonian) are linear combinations
of the generators $S_1,S_2,\cdots,S_n$ of a finite Lie algebra.
This knowledge has been reviewed in the first part of the present work
\cite {KN08}, hereafter called Part I.
In addition, a method was presented to obtain expressions for $U(t)$
in a systematic manner. The aim of what follows is to show that
the method is capable of reproducing known results and of
obtaining new results.

%survey of the paper
Sections 2 and 3 of the paper deal with the SU(2) symmetry.
Section 2 discusses the Bloch equations and a generalisation involving Jacobi's
elliptic functions. Section 3 discusses the case when phase modulation is included.
Section 4 deals with SU(1,1) symmetry, more specifically, the harmonic oscillator
with time-dependent frequency. Section 5 considers more general oscillators.
Finally, Section 6 contains a short discussion of the obtained results
and of the possibilities for further work.

%%%%%%%%%%%%%%%%%%%%%%%%%%%%%%%%%%%%%%%%%%%%%%%%%%%%%%%%%%%%%%%%%%%%%%%%%%%%%%%%%
\section{SU(2)}
\label {case2}

The generators of SU(2) satisfy the commutation relations
\be
[S_1,S_2]=iS_3\quad\mbox{ and cyclic permutations.}
\ee
A well-known application of this Lie algebra concerns the Bloch
equations --- see \cite {AE87}.
It is treated in the present Section. The modification
obtained by considering phase modulation is treated
in the next Section. The combination of
amplitude and of phase modulation was considered in \cite {HHW88}
but will not be considered here.

Note that the incorporation of the Bloch equations
into the Maxwell-Bloch equations have been studied by many authors,
including \cite {AS92, SP06}.

More general applications of SU(2) symmetry have been considered in the literature as well.
Campolieti and Sanctuary \cite {CS89} applied the Wei-Norman technique to
field modulation in NMR. 
Zhou and Ye \cite {ZY94} study the case where all coefficients are time-dependent.
They introduce Euler angles with the same purpose as in the present work.
Finally, Dasgupta \cite {DA98} studies the Jaynes-Cummings model with time-dependent coupling
between the spin and the photon field.

%%%%%%%%%%%%%%%%%%%%%%%%%%%%%%%%%%%%%%%%%%%%%%%%%%%%%%%%%%%%%%%%%%%%%%%%%%%%%%%%%
\subsection{The Bloch equations}

A magnetic spin in a magnetic field is usually described by a Hamiltonian of the form
\be
H=\frac 12\epsilon\sigma_3-\frac 12\xi\sigma_1.
\label {s2:ham}
\ee
The Pauli matrices $\sigma_1,\sigma_2,\sigma_3$ are related to the generators
of the Lie algebra by $\sigma_\alpha=2S_\alpha$.
One has $h=(-\xi,0,\epsilon)^{\rm T}$. The equations of motion $\dot a=h\times a$
are known as the Bloch equations --- see \cite {AE87}.
Written explicitly, they are
\be
\dot a_1&=&-\epsilon a_2,\crcr
\dot a_2&=&\epsilon a_1+\xi a_3\crcr
\dot a_3&=&-\xi a_2.
\label {su2:emeq}
\ee

%%%%%%%%%%%%%%%%%%%%%%%%%%%%%%%%%%%%%%%%%%%%%%%%%%%%%%%%%%%%%%%%%%%%%%%%%%%%%%%%%
\subsection{Special solutions}

Assume that the constant $\epsilon$ does not depend on time.
If $\xi(t)$ does not depend on time
then any solution $x(t)$ of the harmonic oscillator
equation $\ddot x+(\epsilon^2+\xi^2)x=0$ determines a solution of
(\ref {su2:emeq}), given by
\be
a_1&=&\frac {\epsilon \dot x}{\epsilon^2+\xi^2}+\xi C\crcr
a_2&=&x,\crcr
a_3&=&\frac {\xi\dot x}{\epsilon^2+\xi^2}-\epsilon C,
\label {su2sol1}
\ee
with integration constant $C$.

When  $\xi(t)$ is time-dependent then a solution is known
only in very specific cases.
One such case involves Jacobi's elliptic functions
$\sn$, $\cn$, and $\dn$, with  elliptic modulus $k$.
Let
\be
\xi(t)=2\omega k\cn(\omega t;k).
\label {s2:xit}
\ee
Then a solution of (\ref {su2:emeq}) exists of the form
\be
a_1(t)&=&\epsilon\cn(\omega t;k),\crcr
a_2(t)&=&\omega \sn(\omega t;k)\dn(\omega t;k),\crcr
a_3(t)&=&-\omega k\sn^2(\omega t;k)+\gamma,
\label {s2:spsol}
\ee
with
\be
\gamma=-\frac {\epsilon^2-\omega^2}{2\omega k}.
\label {s2:gamma}
\ee
In the limit $k=1$ these expressions lead to the well-known result
(see Eq.~4.21 of \cite {AE87})
\be
\xi(t)&=&2\omega\sech(\omega t),\crcr
a_1(t)&=&\epsilon\sech(\omega t),\crcr
a_2(t)&=&\omega\tanh(\omega t)\sech(\omega t),\crcr
a_3(t)&=&-\omega\tanh^2(\omega t)+\gamma.
\ee
When $\omega=\epsilon$ then the limit $k=0$ can be taken. The rather trivial result is
\be
\xi(t)&=&0,\crcr
a_1(t)&=&\omega\cos(\omega t),\crcr
a_2(t)&=&\omega \sin(\omega t).\crcr
a_3(t)&=&0.
\ee

%%%%%%%%%%%%%%%%%%%%%%%%%%%%%%%%%%%%%%%%%%%%%%%%%%%%%%%%%%%%%%%%%%%%%%%%%%%%%%%%%
\subsection{Automorphisms}

Consider the special solution (\ref {s2:spsol}).
Following the general method of Section 4.4 of Part I
the transformation $V(t)$ is determined by two angles
$\phi(t),\theta(t)$. The special solution $a(t)$
at time $t=0$ reads
\be
a(0)=\left(\epsilon,0,\gamma\right)^{\rm T}.
\ee
It is rotated into the fixed vector
$\lambda(0,1,0)^{\rm T}$.
One has
\be
z(t)&=&\sqrt{a_1^2(t)+a_2^2(t)}=\sqrt{\epsilon^2\cn^2(\omega t;k)+\omega^2\sn^2(\omega t;k)\dn^2(\omega t;k)},\crcr
z(0)&=&\epsilon,\crcr
\lambda&=&\sqrt{z^2(t)+a_3^2(t)}=\sqrt{\epsilon^2+\gamma^2}.
\label {s2:zlambda}
\ee
The angles $\phi(t)$ and $\theta(t)$ are determined by (62) of Part I.
In particular, at $t=0$ is
\be
\sin(\phi(0))=\frac {a_1(0)}{z(0)}=1&\quad&\mbox { and }\quad
\cos(\phi(0))=\frac {a_2(0)}{z(0)}=0,\crcr
\sin(\theta(0))=-\frac {a_3(0)}\lambda=-\frac{\gamma}{\lambda}&\quad&\mbox{ and }\quad
\cos(\theta(0))=\frac {z(0)}{\lambda}=\frac{\epsilon}{\lambda}.
\ee
One can understand these values as follows.
By a rotation of $-\phi(0)=-\pi/2$ around the third axis the initial vector $a(0)$
becomes $(0,\epsilon,\gamma)^{\rm T}$. Then by a rotation with angle $-\theta(0)$
around the first axis it becomes $(0,\lambda,0)^{\rm T}$.
Next, the rotation $R_1(\theta(t))$, followed by the rotation $R_3(\phi(t))$
maps this fixed vector onto the time-dependent $a(t)$.

The Hamiltonian $K(t)$ equals (see (63) of Part I)
\be
K(t)
&=&\frac {a_2\dot a_3}{z^2}S_1-\frac {a_1\dot a_3}{z^2}S_2+\frac {a_1\dot a_2-a_2\dot a_1}{z^2}S_3\crcr
&=&\frac{\xi(t)}{z^2(t)}\left[-a_2^2(t)S_1+a_1(t)a_2(t)S_2+a_1(t)a_3(t)S_3\right]
+\epsilon S_3.
\label {su2:k}
\ee
The difference between this $K(t)$ and the Hamiltonian $H(t)$ as given by (\ref {s2:ham})
makes an extra rotation necessary. It involves the function $\alpha(t)$, given by
(64) of Part I. It evaluates to
\be
\alpha(t)
&=&-\frac {a_1(t)}{z^2(t)}\xi(t)
=-\frac {2\epsilon\omega k\cn^2(\omega t;k)}{\epsilon^2\cn^2(\omega t;k)+\omega^2\sn^2(\omega t;k)\dn^2(\omega t;k)}.
\label {s2:alpha}
\ee
The final result then becomes
\be
U(t)=e^{i\phi(t)S_3}e^{i(\theta(t)-\theta(0))S_1}e^{-i(\pi/2)S_3}
e^{-i\left(\int_0^t{\rm d}s\,\alpha(s)\right)(\epsilon S_1+\gamma S_3)}.
\ee
Note that this expression can be simplified to
\be
U(t)=e^{i(\phi(t)-\pi/2)S_3}e^{-i(\theta(t)-\theta(0))S_2}
e^{i\lambda \tau X}.
\label {su2:finres}
\ee
with $\displaystyle \tau\equiv\tau(t)=-\int_0^t{\rm d}s\,\alpha(s)$
and $\displaystyle X=\frac {\epsilon S_1+\gamma S_3}\lambda$.
For further use note that
\be
e^{i\lambda\tau X}S_j e^{-i\lambda\tau X}=S_j+i\sin(\lambda\tau)[X,S_j]
+(\cos(\lambda\tau)-1)[X,[X,S_j]].
\ee

%%%%%%%%%%%%%%%%%%%%%%%%%%%%%%%%%%%%%%%%%%%%%%%%%%%%%%%%%%%%%%%%%%%%%%%%%%%%%%%%%
\subsection {General solution of the Bloch equations}

The general solution of the Bloch equations, given arbitrary initial conditions
and with time-dependent Hamiltonian determined by (\ref {s2:xit}),
can be derived using (\ref {su2:finres}). 
Note that (omitting time dependences and denoting $\theta_0\equiv\theta(0)$)
\be
\epsilon\cos(\theta-\theta_0)-\gamma\sin(\theta-\theta_0)
&=&\frac {\epsilon^2-\gamma^2}\lambda\cos(\theta)-2\frac {\epsilon\gamma}\lambda\sin(\theta)\crcr
\gamma\cos(\theta-\theta_0)+\epsilon\sin(\theta-\theta_0)
&=&2\frac {\epsilon\gamma}\lambda\cos(\theta)+\frac {\epsilon^2-\gamma^2}\lambda\sin(\theta)\crcr
\gamma\cos(\theta-\theta_0)-\epsilon\sin(\theta-\theta_0)
&=&-\lambda\sin(\theta)\crcr
\epsilon\cos(\theta-\theta_0)+\gamma\sin(\theta-\theta_0)
&=&\lambda\cos(\theta).
\ee

These relations can be used to calculate
\be
U(t)(\epsilon S_1+\gamma S_3)U(t)^*
&=&-\lambda\sin(\theta)S_3+\lambda\cos(\theta)S_+,\crcr
%%%%%%
\lambda U(t)(\epsilon S_1-\gamma S_3)U(t)^*
&=&\left[(\epsilon^2-\gamma^2)\cos(\theta)-2\epsilon\gamma\sin(\theta)\cos(\lambda\tau)\right]
S_+\crcr
& &
+2\gamma\epsilon \sin(\lambda\tau)S_-\crcr
& &-\left[2\epsilon\gamma\cos(\theta)\cos(\lambda\tau)+(\epsilon^2-\gamma^2)\sin(\theta)\right]
S_3\crcr
%%%%%%
U(t)S_2U(t)^*
&=&-\sin(\lambda\tau)\cos(\theta)S_3
-\cos(\lambda\tau)S_-
-\sin(\theta)\sin(\lambda\tau)S_+,\crcr
& &
\label {max:timev}
\ee
with
\be
S_+=\sin(\phi)S_1+\cos(\phi)S_2
\quad\mbox{ and }
S_-=\cos(\phi)S_1-\sin(\phi)S_2.
\ee
Note further that
\be
\dot\phi&=&-\epsilon+\xi\sin(\phi)\tan(\theta)\cr
\dot\theta&=&\xi\cos(\phi)\cr
\dot\tau&=&\frac \xi\lambda\,\frac {\sin(\phi)}{\cos(\theta)}.
\ee
Using these equations one can verify explicitly that the time-dependent operators (\ref {max:timev})
satisfy indeed the von Neumann equation of motion.

From (\ref {max:timev}) one can obtain the three independent solutions
$a^{(1)}, a^{(2)},a^{(3)}$ of the Bloch equations.
\be
a^{(1)}
&=&\frac \epsilon{\lambda}
\left(\begin{array}{c}
       \sin(\phi)\cos(\theta)\\\cos(\phi)\cos(\theta)\\-\sin(\theta)
      \end{array}\right)
-\frac\gamma\lambda\cos(\lambda\tau)
\left(\begin{array}{c}
       \sin(\phi)\sin(\theta)\\\cos(\phi)\sin(\theta)\\\cos(\theta)
      \end{array}\right)
+\frac\gamma\lambda\sin(\lambda\tau)
\left(\begin{array}{c}
       \cos(\phi)\\-\sin(\phi)\\0
      \end{array}\right)\crcr
%%%%%%%%%%%%%%%%%%%
a^{(2)}&=&-\sin(\theta)\sin(\lambda\tau)\left(\begin{array}{c}
       \sin(\phi)\\\cos(\phi)\\0
      \end{array}\right)
-\cos(\lambda\tau)\left(\begin{array}{c}
       \cos(\phi)\\-\sin(\phi)\\0
      \end{array}\right)
-\cos(\theta)\sin(\lambda\tau)\left(\begin{array}{c}
       0\\0\\1
      \end{array}\right)\crcr
%%%%%%%%%%%%%%%%%%%
a^{(3)}
&=&\frac \gamma{\lambda}
\left(\begin{array}{c}
       \sin(\phi)\cos(\theta)\\\cos(\phi)\cos(\theta)\\-\sin(\theta)
      \end{array}\right)
+\frac \epsilon{\lambda}\cos(\lambda\tau)
\left(\begin{array}{c}
       \sin(\phi)\sin(\theta)\\\cos(\phi)\sin(\theta)\\\cos(\theta)
      \end{array}\right)
-\frac\epsilon\lambda\sin(\lambda\tau)
\left(\begin{array}{c}
       \cos(\phi)\\-\sin(\phi)\\0
      \end{array}\right)\crcr
& &
\ee
The special solution (\ref {s2:spsol}) satisfies $a=\epsilon a^{(1)}+\gamma a^{(3)}$.
The general solution of the Bloch equations $\dot x=h\times x$
is given by expression (67) of Part I and agrees with the results given above.

%%%%%%%%%%%%%%%%%%%%%%%%%%%%%%%%%%%%%%%%%%%%%%%%%%%%%%%%%%%%%%%%%%%%%%%%%%%%%%%%%
\subsection {The $k=1$-limit}

We now focus on the case $k=1$ because then the formulas simplify.
The relevant expressions become
\be
\sin(\phi)&=&\frac {a_1}z=\frac {\epsilon}{\sqrt{\epsilon^2+\omega^2\tanh^2(\omega t)}},\crcr
\cos(\phi)&=&\frac {a_2}z=\frac {\omega \tanh(\omega t)}{\sqrt{\epsilon^2+\omega^2\tanh^2(\omega t)}},\crcr
\sin(\theta)&=&-\frac {a_3}\lambda=\frac {\omega\tanh^2(\omega t)-\gamma}{\sqrt{\epsilon^2+\gamma^2}},\crcr
\cos(\theta)&=&\frac z\lambda=\sech(\omega t)\sqrt{\frac {\epsilon^2+\omega^2\tanh^2(\omega t)}{\epsilon^2+\gamma^2}},\crcr
z&=&\sqrt{a_1^2+a_2^2}=\sech(\omega t)\sqrt{\epsilon^2+\omega^2\tanh(\omega t)},\crcr
\lambda&=&\sqrt{z^2+a_3^2}=\sqrt{\epsilon^2+\gamma^2},\crcr
\alpha&=&-\frac {a_1}{z^2}\xi=-\frac {2\epsilon\omega}{\epsilon^2+\omega^2\tanh^2(\omega t)},
\ee
with $\gamma=(\omega^2-\epsilon^2)/2\omega$.
The integral of $\alpha(t)$ can be done analytically. It yields
\be
\tau(t)=-\int_0^t{\rm d}s\,\alpha(s)=\frac {2\omega}{\epsilon^2+\omega^2}
\left\{\epsilon t+\arctan\left(\frac\omega\epsilon\tanh(\omega t)\right)\right\}.
\ee

%%%%%%%%%%%%%%%%%%%%%%%%%%%%%%%%%%%%%%%%%%%%%%%%%%%%%%%%%%%%%%%%%%%%%%%%%%%%%%%%%
\subsection {The Bloch equations at resonance}

The resonant condition is $\epsilon=\omega$.
In that case, $\gamma$, as given by (\ref {s2:gamma}), vanishes.
Hence, the normalisations $z(t)$ and $\lambda$, given by (\ref {s2:zlambda}) and
appearing in the expressions for the angles $\phi$ and $\theta$
(see (62) of Part I), become
\be
z(t)&=&\omega \sqrt{\cn^2(\omega t;k) + \sn^2(\omega t;k)\dn^2(\omega t;k)},\\
\lambda &=&\omega.
\ee
There follows
\be
\sin(2\phi)
&=&\frac {2a_1a_2}{z^2}=\sn(2\omega t;k)\\
\cos(2\phi)
&=&\frac {a_2^2-a_1^2}{z^2}
=-\cn(2\omega t;k)\\
\sin(\theta)
&=&-\frac {a_3}{\lambda}
=k\sn^2(\omega t;k),\\
\cos(\theta)
&=&\frac z\lambda
=\sqrt{\cn^2(\omega t;k)
 + \sn^2(\omega t;k)\dn^2(\omega t;k)}.
\ee
In particular is $\theta(0)=0$.
The equations (\ref {max:timev}) become
\be
U(t)S_1U(t)^*
&=&-\sin(\theta)S_3+\cos(\theta)S_+,\crcr
%%%%%%
U(t)S_3U(t)^*
&=&\sin(\theta)\cos(\omega\tau)S_+-\sin(\omega\tau)S_-
+\cos(\omega\tau)\cos(\theta)S_3\crcr
%%%%%%
U(t)S_2U(t)^*
&=&-\sin(\omega\tau)\cos(\theta)S_3
-\cos(\omega\tau)S_-
-\sin(\theta)\sin(\omega\tau)S_+.
\ee
They can be used to obtain the general solution of the Bloch equations (\ref {su2:emeq}) at resonance
\be
a^{\rm gen}_1(t) &=&a^{\rm gen}_1(0)\cos(\theta)\sin(\phi)\crcr
& &+\sin(\theta)\sin(\phi)\left[a^{\rm gen}_3(0)\cos(\omega \tau)-a^{\rm gen}_2(0)\sin(\omega\tau)\right]\crcr
& &-\cos(\phi)\left[a^{\rm gen}_2(0)\cos(\omega\tau)+a^{\rm gen}_3(0)\sin(\omega \tau)\right]\crcr
a^{\rm gen}_2(t) &=& a^{\rm gen}_1(0)\cos(\theta)\cos(\phi)\crcr
& &+\sin(\theta)\cos(\phi)\left[a^{\rm gen}_3(0)\cos(\omega \tau)-a^{\rm gen}_2(0)\sin(\omega\tau)\right]\crcr
& &+\sin(\phi)\left[a^{\rm gen}_2(0)\cos(\omega\tau)+a^{\rm gen}_3(0)\sin(\omega \tau)\right]\crcr
a^{\rm gen}_3(t) &=&-a_1^{\rm gen}(0)\sin(\theta)\crcr
& &+ \left[a^{\rm gen}_3(0)\cos(\omega \tau)-a^{\rm gen}_2(0)\sin(\omega\tau)\right]
\cos(\theta).
\label {s2resonancegensol}
\ee

The correction angle $\alpha$, given by (\ref {s2:alpha}), simplifies to
\be
\alpha(t)&=&-\frac {a_1}{z^2}\xi
=-k\omega\left(1+cn(2\omega t;k)\right).
\ee
This expression can be integrated analytically. The result is
\be
\tau(t)=-\int_0^t{\rm d}s\,\alpha(s)=kt-\frac {1}{\omega}
\mbox{ arctan }\frac {\dn(\omega t;k)}{k\sn(\omega t;k)\cn(\omega t;k)}.
\ee
With some effort, (\ref {s2resonancegensol}) can now be written as
\be
a^{\rm gen}_1(t) &=& a^{\rm gen}_1(0)\cn(\omega t;k)\crcr
& & + a^{\rm gen}_2(0)\sn(\omega t;k)\cos(k\omega t)\crcr
& & + a^{\rm gen}_3(0)\sn(\omega t;k)\sin(k\omega t),\crcr
a^{\rm gen}_2(t) &=& a^{\rm gen}_1(0)\sn(\omega t;k)\dn(\omega t;k)\crcr
& & + a^{\rm gen}_2(0)\left(\cn(\omega t;k)\dn(\omega t;k)\cos(k\omega t)
 - k\sn(\omega t;k)\sin(k\omega t)\right) \crcr 
&+& a^{\rm gen}_3(0)\left(\cn(\omega t;k)\dn(\omega t;k)\sin(k\omega t)
 + k\sn(\omega t;k)\cos(k\omega t)\right),\crcr
a^{\rm gen}_3(t) &=&-a_1^{\rm gen}(0)k\sn^2(\omega t;k) \crcr
& &+ a^{\rm gen}_2(0)\left(k\cn(\omega t;k)\sn(\omega t;k)\cos(k\omega t)
 + \dn(\omega t;k)\sin(k\omega t)\right) \crcr 
&+& a^{\rm gen}_3(0)\left(k\cn(\omega t;k)\sn(\omega t;k)\sin(k\omega t)
 -  \dn(\omega t;k)\cos(k\omega t)\right).
\ee
The three independent solutions are therefore
\be
a^{(1)}&=&\left(\begin{array}{c}
               \cn(\omega t;k)\\\sn(\omega t;k)\dn(\omega t;k)\\-k\sn^2(\omega t;k)
              \end{array}\right),\crcr
a^{(2)}&=&\left(\begin{array}{c}
               \sn(\omega t;k)\cos(k\omega t)\\
               \cn(\omega t;k)\dn(\omega t;k)\cos(k\omega t) - k\sn(\omega t;k)\sin(k\omega t)\\
		k\cn(\omega t;k)\sn(\omega t;k)\cos(k\omega t)+ \dn(\omega t;k)\sin(k\omega t)
              \end{array}\right),\crcr
a^{(3)}&=&\left(\begin{array}{c}
               \sn(\omega t;k)\sin(k\omega t)\\
		\cn(\omega t;k)\dn(\omega t;k)\sin(k\omega t) + k\sn(\omega t;k)\cos(k\omega t)\\
		k\cn(\omega t;k)\sn(\omega t;k)\sin(k\omega t) -  \dn(\omega t;k)\cos(k\omega t)
              \end{array}\right).
\ee
One clearly has $a=\omega a^{(1)}$, the special solution we started with.
It is easy to verify that also $a^{(2)}$ and $a^{(3)}$ are solutions of
$\dot a=a\times h$.

%%%%%%%%%%%%%%%%%%%%%%%%%%%%%%%%%%%%%%%%%%%%%%%%%%%%%%%%%%%%%%%%%%%%%%%%%%%%%%%%%
\section{SU(2) continued}
\label {case1}

%%%%%%%%%%%%%%%%%%%%%%%%%%%%%%%%%%%%%%%%%%%%%%%%%%%%%%%%%%%%%%%%%%%%%%%%%%%%%%%%%
\subsection{Including phase modulation}

A slightly different solution is obtained when
the parameter $\epsilon$ in (\ref {su2:emeq})
is made time dependent in the following way
\be
\epsilon=\epsilon_0\tanh(\omega t).
\label {phasemod:epstd}
\ee
In \cite {AE87}, the resulting equations are called the Bloch equations
including phase modulation. Also in this case a special solution is known.
The generalisation to Jacobi's elliptic functions, as given below, can be done in two different ways.

%%%%%%%%%%%%%%%%%%%%%%%%%%%%%%%%%%%%%%%%%%%%%%%%%%%%%%%%%%%%%%%%%%%%%%%%%%%%%%%%%
\subsection{Special solutions}

Assume that
\be
\xi(t)&=&\xi_0\cn(\omega t;k),\crcr
\epsilon(t)&=&\epsilon_0\sn(\omega t;k).
\label {phasemod:case1}
\ee
or
\be
\xi(t)&=&\xi_0\dn(\omega t;k),\crcr
\epsilon(t)&=&\epsilon_0\sn(\omega t;k).
\label {phasemod:case2}
\ee
Both assumptions reduce to $\xi(t)=\xi_0\sech(\omega t)$ and (\ref {phasemod:epstd})
in the limit $k=1$.

A solution of the equations (\ref {su2:emeq}) is given by
\be
a_1(t)&=&\epsilon_0\cn(\omega t;k),\crcr
a_2(t)&=&\omega\dn(\omega t;k),\crcr
a_3(t)&=&-\xi_0\sn(\omega t; k).
\label {s22:spsol}
\ee
or
\be
a_1(t)&=&\omega\cn(\omega t;k),\crcr
a_2(t)&=&\epsilon_0\dn(\omega t;k),\crcr
a_3(t)&=&-\xi_0\sn(\omega t; k).
\label {s22:spsolbis}
\ee
provided that $\xi_0^2=\epsilon_0^2+\omega^2k^2$,
respectively $\xi_0^2=\epsilon_0^2k^2+\omega^2$, is satisfied.

Only the first of the cases is treated below. The other case is completely analogous.

Note that in the limit $k=1$ the solution (\ref {s22:spsol}) reduces to the well-known solution
\be
a_1(t)&=&\epsilon_0\sech(\omega t),\crcr
a_2(t)&=&\omega\sech(\omega t),\crcr
a_3(t)&=&-\xi_0\tanh(\omega t).
\ee
In the limit $k=0$ it reduces to a harmonic precession
\be
\xi(t)&=&\xi_0\cos(\omega t),\crcr
\epsilon(t)&=&\epsilon_0\sin(\omega t),\crcr
a_1(t)&=&\epsilon_0\cos(\omega t),\crcr
a_2(t)&=&\omega,\crcr
a_3(t)&=&-\xi_0\sin(\omega t).
\ee

%%%%%%%%%%%%%%%%%%%%%%%%%%%%%%%%%%%%%%%%%%%%%%%%%%%%%%%%%%%%%%%%%%%%%%%%%%%%%%%%%
\subsection{Automorphisms}

Consider the special solution (\ref {s22:spsol}).
The angles $\phi(t)$ and $\theta(t)$ are determined by (62) of Part I.
In particular, at $t=0$ is
\be
\sin(\phi(0))=\frac {\epsilon_0}{\lambda},
\quad
\cos(\phi(0))=\frac {\omega}{\lambda},
\ee
and $\theta(0)=0$, with $\lambda=\sqrt {\epsilon_0^2+\omega^2}$.
This means that by a rotation $R_3(-\phi(0))$ around the third axis the
initial vector $a(0)$ is rotated into the fixed vector $\lambda(0,1,0)^{\rm T}$.
Next, the rotation $R_1(\theta(t))$, followed by the rotation $R_3(\phi(t))$
maps this fixed vector onto the time-dependent $a(t)$.

The Hamiltonian $K(t)$ equals (see (63) of Part I)
\be
K(t)
&=&\frac {a_2\dot a_3}{z^2}S_1-\frac {a_1\dot a_3}{z^2}S_2+\frac {a_1\dot a_2-a_2\dot a_1}{z^2}S_3\crcr
&=&\frac{\xi(t)}{z^2(t)}\left[-a_2^2(t)S_1+a_1(t)a_2(t)S_2+a_1(t)a_3(t)S_3\right]
+\epsilon S_3.
\ee
This is the same expression as (\ref {su2:k}).
The difference between this $K(t)$ and the Hamiltonian $H(t)$ as given by (\ref {s2:ham})
makes an extra rotation necessary. It involves the function $\alpha(t)$, given by
(64) of Part I. It evaluates to
\be
\alpha(t)
&=&-\frac {a_1(t)}{z^2(t)}\xi(t)
=-\frac {\epsilon_0\xi_0\cn(\omega t;k)\dn(\omega t;k)}{\epsilon_0^2\cn^2(\omega t;k)+\omega^2\dn^2(\omega t;k)}.
\label {phasemod:alpha1}
\ee
The final result then becomes
\be
U(t)=e^{i\phi(t)S_3}e^{i\theta(t)S_1}e^{-i\phi(0)S_3}
e^{i\lambda\tau(t) X},
\ee
with $\displaystyle \tau(t)=-\int_0^t{\rm d}s\,\alpha(s)$
and $\displaystyle X=\frac {\epsilon_0 S_1+\omega S_2}{\lambda}$.
Note that (\ref {phasemod:alpha1}) can be integrated analytically. The result is
\be
\tau(t)=\frac{\epsilon_0\xi_0}{2\omega\lambda\mu}\ln\frac {\lambda+\mu\sn(\omega t;k)}{\lambda-\mu\sn(\omega t;k)}
\ee
with $\displaystyle\mu=\sqrt{\epsilon_0^2+k^2\omega^2}$.

%%%%%%%%%%%%%%%%%%%%%%%%%%%%%%%%%%%%%%%%%%%%%%%%%%%%%%%%%%%%%%%%%%%%%%%%%%%%%%%%%
\subsection{General solution including phase modulation}

It is now possible to obtain the general solution of the Bloch
equations including phase modulation, with driving fields of the form
(\ref {phasemod:case1}).
One calculates (again omitting time dependencies and denoting $\phi_0\equiv\phi(0)$)
\be
U(t)(\epsilon_0 S_1+\omega S_2)U(t)^*
&=&\lambda\cos(\theta)S_+
-\lambda\sin(\theta)S_3\crcr
%%%%%%%%%%%%%%
\lambda U(t)(\epsilon_0 S_1-\omega S_2)U(t)^*&=&
\left[(\epsilon_0^2-\omega^2)\cos(\theta)
+2\epsilon_0\omega\sin(\theta)\sin(\lambda\tau)\right]S_+\crcr
& &+2\epsilon_0\omega\cos(\lambda\tau)S_-\crcr
& &+\left[2\epsilon_0\omega\cos(\theta)\sin(\lambda\tau)
-(\epsilon_0^2-\omega^2)\sin(\theta)\right]S_3\crcr
%%%%%%%%%%%%
U(t)S_3U(t)^*
&=&\cos(\lambda\tau)\sin(\theta)S_+
-\sin(\lambda\tau)S_-
+\cos(\lambda\tau)\cos(\theta)S_3,
\ee
with
\be
S_+=\sin(\phi)S_1+\cos(\phi)S_2
\quad\mbox{ and }
S_-=\cos(\phi)S_1-\sin(\phi)S_2.
\ee
The three independent solutions of the Bloch
equations including phase modulation are therefore
\be
a^{(1)}(t)&=&\frac{\epsilon_0}\lambda
\left(\begin{array}{c}
       \sin(\phi)\cos(\theta)\\\cos(\phi)\cos(\theta)\\-\sin(\theta)
      \end{array}\right)\crcr
& &
+\frac\omega\lambda\sin(\lambda\tau)
\left(\begin{array}{c}
       \sin(\phi)\sin(\theta)\\\cos(\phi)\sin(\theta)\\\cos(\theta)
      \end{array}\right)
+\frac\omega\lambda\cos(\lambda\tau)
\left(\begin{array}{c}
       \cos(\phi)\\-\sin(\phi)\\0
      \end{array}\right),\cr
%%%%%%%%%%%%%%%%%%
a^{(2)}(t)&=&\frac\omega\lambda
\left(\begin{array}{c}
       \sin(\phi)\cos(\theta)\\\cos(\phi)\cos(\theta)\\-\sin(\theta)
      \end{array}\right)\crcr
& &
-\frac{\epsilon_0}\lambda\sin(\lambda\tau)
\left(\begin{array}{c}
       \sin(\phi)\sin(\theta)\\\cos(\phi)\sin(\theta)\\ \cos(\theta)
      \end{array}\right)
-\frac{\epsilon_0}\lambda\cos(\lambda\tau)
\left(\begin{array}{c}
       \cos(\phi)\\-\sin(\phi)\\0
      \end{array}\right),\cr
%%%%%%%%%%%%%%%%%%%%%
a^{(3)}(t)&=&\cos(\lambda\tau)
\left(\begin{array}{c}
       \sin(\phi)\sin(\theta)\\\cos(\phi)\sin(\theta)\\\cos(\theta)
      \end{array}\right)
-\sin(\lambda\tau)\left(\begin{array}{c}
       \cos(\phi)\\-\sin(\phi)\\0
      \end{array}\right).
\ee
The special solution (\ref {s22:spsol}) satisfies $a=\epsilon_0 a^{(1)}+ \omega a^{(2)}$.
The general solution of the equations $\dot x=h\times x$ has been given in Part I of the paper
and can be rederived from the knowledge of $a^{(1)},a^{(2)},a^{(3)}$.

%%%%%%%%%%%%%%%%%%%%%%%%%%%%%%%%%%%%%%%%%%%%%%%%%%%%%%%%%%%%%%%%%%%%%%%%%%%%%%%%%
\subsection {The $k=1$-limit}

In this limit the existence of the special solution requires that $\xi_0=\lambda$.
The angle $\phi$ and the phase $\alpha$ become constants
\be
\sin(\phi)=\frac {\epsilon_0}\lambda
\quad\mbox{ and }\quad
\cos(\phi)=\frac{\omega}{\lambda}
\ee
and $\alpha=-\epsilon_0\xi_0/\lambda^2$. Hence one has
$\lambda\tau(t)=\epsilon_0 t$.
The angle $\theta$ satisfies
\be
\sin(\theta)=-\tanh(\omega t)
\quad\mbox{ and }\quad
\cos(\theta)=\sech(\omega t).
\ee
The three independent solutions are
\be
a^{(1)}(t)&=&\frac {\epsilon_0}{\lambda^2}
\left(\begin{array}{c}
       \epsilon_0\sech(\omega t)\\\omega \sech(\omega t)\\\lambda\tanh(\omega t)
      \end{array}\right)
+\frac{\omega}{\lambda^2}\sin(\epsilon_0 t)
\left(\begin{array}{c}
       -\epsilon_0\tanh(\omega t)\\-\omega \tanh(\omega t)\\\lambda\sech(\omega t)
      \end{array}\right)
+\frac{\omega}{\lambda^2}\cos(\epsilon_0 t)
\left(\begin{array}{c}
       \omega\\-\epsilon_0\\0
      \end{array}\right)\crcr
a^{(2)}(t)&=&\frac {\omega}{\lambda^2}
\left(\begin{array}{c}
       \epsilon_0\sech(\omega t)\\\omega \sech(\omega t)\\\lambda\tanh(\omega t)
      \end{array}\right)
-\frac{\epsilon_0}{\lambda^2}\sin(\epsilon_0 t)
\left(\begin{array}{c}
       -\epsilon_0\tanh(\omega t)\\-\omega \tanh(\omega t)\\\lambda\sech(\omega t)
      \end{array}\right)
-\frac{\epsilon_0}{\lambda^2}\cos(\epsilon_0 t)
\left(\begin{array}{c}
       \omega\\-\epsilon_0\\0
      \end{array}\right)\crcr
a^{(3)}(t)&=&\frac 1\lambda\cos(\epsilon_0 t)
\left(\begin{array}{c}
       -\epsilon_0\tanh(\omega t)\\-\omega\tanh(\omega t)\\\lambda\sech(\omega t)
      \end{array}\right)
-\frac 1\lambda\sin(\epsilon_0 t)
\left(\begin{array}{c}
       \omega\\-\epsilon_0\\0
      \end{array}\right).
\ee

%%%%%%%%%%%%%%%%%%%%%%%%%%%%%%%%%%%%%%%%%%%%%%%%%%%%%%%%%%%%%%%%%%%%%%%%%%%%%%%%%
\section{SU(1,1)}

The generators of SU(1,1) satisfy the commutation relations
\be
\strut[S_1,S_2]&=&iS_3,\crcr
\strut[S_2,S_3]&=&-iS_1,\crcr
\strut[S_3,S_1]&=&-iS_2.
\label {su11:ccr}
\ee
Consider creation and annihilation operators $b^\dagger$ and $b$
satisfying the canonical commutation relations
$[b,b^\dagger]=\Io$.
Then the operators
\be
S_1&=&\frac 14((b^\dagger)^2+b^2),\crcr
S_2&=&\frac i4((b^\dagger)^2-b^2),\crcr
S_3&=&\frac 14(b^\dagger b+bb^\dagger)
\ee
satisfy (\ref {su11:ccr}). Hamiltonians which can be written
as a linear combination of these generators are the generalised
harmonic oscillators.

%%%%%%%%%%%%%%%%%%%%%%%%%%%%%%%%%%%%%%%%%%%%%%%%%%%%%%%%%%%%%%%%%%%%%%%%%%%%%%%%%
\subsection{Time-dependent frequency}

Consider the harmonic oscillator with arbitrary time-depen\-dent frequency
$\omega(t)$
\be
H=\frac 1{2m}P^2+\frac 12m\omega^2(t)Q^2.
\ee
Here, $Q$ is the position operator and $P$ the momentum operator.
They satisfy $[Q,P]=i$.
This problem was studied by Lewis and Riesenfeld \cite {LR69}.
See also \cite {EVP80,DPB92,YKUGP94,SDY00,HG01,KSP03,GMF04,LW07,DL08}.

Introduce an annihilation operator defined by
\be
b=\frac{1}{r\sqrt 2}Q+i\frac r{\sqrt 2}P,
\ee
with $\displaystyle r=\frac {1}{\sqrt{m\omega_0}}$
and $\omega_0$ some constant frequency.
Then the generators equal
\be
S_1&=&\frac 1{4r^2}Q^2-\frac {r^2}4P^2\\
S_2&=&\frac 14(QP+PQ)\\
S_3&=&\frac 1{4r^2}Q^2+\frac {r^2}4P^2.
\ee
Introduce the function $\gamma(t)$, modulating the frequency $\omega_0$, defined by
$\displaystyle \omega(t)=\gamma(t)\omega_0$.
The Hamiltonian becomes
\be
H(t)
&=&-\frac 14\omega_0(b-b^\dagger)^2+\frac 14\omega_0\gamma^2(t)(b+b^\dagger)^2\crcr
&=&\omega_0(\gamma^2(t)-1)S_1+\omega_0(\gamma^2(t)+1)S_3.
\ee
Hence one has $\displaystyle h=\omega_0(\gamma^2(t)-1,0,\gamma^2(t)+1)^{\rm T}$.

%%%%%%%%%%%%%%%%%%%%%%%%%%%%%%%%%%%%%%%%%%%%%%%%%%%%%%%%%%%%%%%%%%%%%%%%%%%%%%%%%
\subsection{Special solution}

The time evolution equation reads
(see the definition of the Lie bracket in Part I, equation (69))
\be
\dot a
&=&h\times a\crcr
&=&(h_3a_2,h_1a_3-h_3a_1,h_1a_2)^{\rm T}\crcr
&=&\omega_0(\gamma^2(t)+1)(a_2,-a_1,0)^{\rm T}
+\omega_0(\gamma^2(t)-1)(0,a_3,a_2)^{\rm T}.
\label {tdf:evolution}
\ee

\begin {proposition}
Let $x(t)$ be a solution of the classical oscillator equation
\be
\ddot x+\omega^2(t)x=0.
\label {tdf:classosc}
\ee
Then $a$, defined by
\be
a=\frac 12(\dot x)^2(1,0,1)^{\rm T}
+\frac 12\omega_0^2x^2(-1,0,1)^{\rm T}-\omega_0 {x\dot x}(0,1,0)^{\rm T},
\label {hospec}
\ee
is a solution of (\ref {tdf:evolution}).
\end {proposition}

The proof is done by explicit calculation.
One concludes that, to find a special solution of the von Neumann equation, it suffices to find
a special solution of the classical equation (\ref {tdf:classosc}).
Note that the latter problem is equivalent with solving
Riccati's equation
\be
\dot g-g^2=\omega^2(t).
\ee
The corresponding solution of (\ref {tdf:classosc}) is
\be
x=C\exp\left(-\int{\rm d}t\,g(t)\right),
\ee
with integration constant $C$.
In the case that $\omega(t)$ is constant one finds
\be
g(t)=\omega\tan(\omega t)
\ee
so that $x(t)=C\cos(\omega t)$. With $C=1$ and $\omega_0=\omega$
one obtains the special solution
\be
a=\frac 12\omega^2(-\cos(2\omega t),\sin (2\omega t),1)^{\rm T}.
\ee
If $\omega(t)$ is of the form $\omega(t)=\omega_0(1+\epsilon\cos(2\lambda t))$
then the equation (\ref {tdf:classosc}) is related to Mathieu's equation.
The solution $x=\cn(\omega_0,t;k)$, involving Jacobi's elliptic function with $0<k<1/\sqrt 2$,
is obtained when
\be
\omega(t)=\omega_0\sqrt{2\dn^2(\omega_0t;k)-1}.
\ee

%%%%%%%%%%%%%%%%%%%%%%%%%%%%%%%%%%%%%%%%%%%%%%%%%%%%%%%%%%%%%%%%%%%%%%%%%%%%%%%%%
\subsection{Automorphisms}

Note that the special solution (\ref {hospec}) satisfies $\langle a|a\rangle=0$
(using the metric with signature $-,-,+$).
Hence, we have to apply the exceptional case discussed in Part I,
at the end of Section 5.

Write $a(t)$ into the form
\be
a(t)=R_3(\phi)P_1(\chi)R_3(-\phi(0))a(0).
\ee
From (77) of Part I follows
\be
\sin(\phi)&=&\frac {a_1}{a_3}=\frac {(\dot x)^2-\omega_0^2x^2}{(\dot x)^2+\omega_0^2x^2},\crcr
\cos(\phi)&=&\frac {a_2}{a_3}=\frac {2\omega_0x\dot x}{(\dot x)^2+\omega_0^2x^2}.
\ee
From (83) of Part I follows
\be
\chi=\ln\frac {a_3(t)}{a_3(0)}=\ln\frac {(\dot x)^2+\omega_0^2x^2\,\,\strut}{[(\dot x)^2+\omega_0^2x^2)]_{t=0}}.
\ee
Note that $\phi(0)$ simplifies if either $x(0)=0$ or $\dot x(0)=0$.

From the general theory now follows that (see (76) of Part I)
\be
V(t)&=&e^{-i\phi(t) S_3}e^{-i\chi(t) S_1}e^{i\phi(0)S_3}.
\ee
The corresponding Hamiltonian is
\be
K=\dot\chi\left\{\cos(\phi)S_1-\sin(\phi)S_2\right\}+\dot\phi S_3.
\ee
Note that
\be
\dot\chi=\frac {\dot a_3}{a_3}=h_1\frac {a_2}{a_3}
\quad\mbox{ and }\quad
\dot\phi=\frac {a_2\dot a_1-a_1\dot a_2}{a_3^2}=h_3-h_1\frac {a_1}{a_3}.
\ee
Hence $K$ can be written as
\be
K&=&(\cos(\phi)\dot\chi,-\sin(\phi)\dot\chi,\dot\phi)^{\rm T}\crcr
&=&\frac 1{a_3^2}\left(h_1a_2^2,-h_1a_1a_2,h_3a_3^2-h_1a_1a_3\right)^{\rm T}.
\ee
One obtains $h-k=\alpha a$, with
\be
\alpha=h_1\frac {a_1}{a_3^2}.
\ee
The final result is
\be
U(t)=V(t)e^{if_t(\rho_s(0)-C)}=e^{-i\phi(t) S_3}e^{-i\chi (t) S_1}e^{i\phi(0)S_3}
e^{i\sum_ja_j(0)S_j\int_0^t{\rm d}s\,\alpha(s)}.
\ee

%%%%%%%%%%%%%%%%%%%%%%%%%%%%%%%%%%%%%%%%%%%%%%%%%%%%%%%%%%%%%%%%%%%%%%%%%%%%%%%%%
\section{More general oscillators}

More general time-dependent harmonic oscillators have been considered
in the literature \cite {GCC85,LCP91,SR91,LKJ96,LL02,CJR03,LW07}.
Even damped oscillators have been studied --- see for instance \cite {TJ93,KSP94,RW02}.
Some of them can be treated by the present method.
An example not yet considered in the literature is
$H=\sum_{j=1}^3h_j(t)S_j$ with
\be
h_1(t)&=&a\cn(\omega t;k)\\
h_2(t)&=&-a\sn(\omega t;k)\\
h_3(t)&=&c+\omega\dn(\omega t;k),
\label {more:ham}
\ee
with constants $\omega$, $a$, $c$, and $0\le k\le 1$.
A special solution of the von Neumann equation is given by (see \cite {NK06})
$\rho_s(t)=\sum_{j=1}^3a_j(t)S_j$ with
\be
a_1(t)&=& a\cn(\omega t;k)\\
a_2(t)&=&-a\sn(\omega t;k)\\
a_3(t)&=&c.
\ee
Note that $\rho_s(t)$ is not a density operator. But this does not
harm our method.

Introduce unitary operators $V(t)$ by (see (76) of Part I)
\be
V(t)=e^{-i\phi(t)S_3}e^{-i\chi(t)S_1}e^{i\chi(0)S_1}e^{i\phi(0)S_3}
\label {more:vt}
\ee
with angles $\phi(t)$ and $\chi(t)$ satisfying (see (77) of Part I)
\be
\sin(\phi)&=& \cn(\omega t;k), \qquad\cos(\phi)=\sn(\omega t;k)\\
\sinh(\chi)&=&\frac {c}{\sqrt{a^2-c^2}}, \qquad \cosh(\chi)=\frac {a}{\sqrt{a^2-c^2}}.
\ee
Note that we assume that $a^2>c^2$.
Also, note that $\phi(0)=\pi/2$.
Because $\chi(t)$ turns out to be independent of time $t$,
(\ref {more:vt}) simplifies to
\be
V(t)=e^{-i(\phi(t)-\phi(0))S_3}.
\label {more:vt2}
\ee
Next calculate
\be
\alpha(t)=\frac {a_1h_1+a_2h_2}{a^2}=1.
\ee
Hence, the time evolution is described by the unitary operators
\be
U(t)=V(t)e^{-i\rho_s(0)t}=e^{-i(\phi(t)-\phi(0))S_3}e^{-it(aS_1+cS_3)}.
\ee
It is now possible to calculate the time evolution of the generators
in the Heisenberg picture. For simplicity take $c=0$. Then one obtains
\be
S_1(t)&=&U(t)^*S_1U(t)\crcr
&=&\cos(\phi(t)-\phi(0))S_1+\sin(\phi(t)-\phi(0))S_2\crcr
&=&\sin(\phi(t))S_1-\cos(\phi(t))S_2\crcr
&=&\cn(\omega t;k)S_1-\sn(\omega t;k)S_2.
\ee
Similarly is
\be
S_2(t)
&=&U(t)^*S_2U(t)\crcr
&=&\cosh(at)\left[\cos(\phi(t)-\phi(0))S_2-\sin(\phi(t)-\phi(0))S_1\right]-\sinh(at)S_3\crcr
&=&\cosh(at)\left[\sin(\phi(t))S_2+\cos(\phi(t))S_1\right]-\sinh(at)S_3\crcr
&=&\cosh(at)\left[\cn(\omega t;k)S_2+\sn(\omega t;k)S_1\right]-\sinh(at)S_3,
\ee
and
\be
S_3(t)
&=&U(t)^*S_3U(t)\crcr
&=&\cosh(at)S_3-\sinh(at)\left[\cos(\phi(t)-\phi(0))S_2-\sin(\phi(t)-\phi(0))S_1\right]\crcr
&=&\cosh(at)S_3-\sinh(at)\left[\sin(\phi(t))S_2+\cos(\phi(t))S_1\right]\crcr
&=&\cosh(at)S_3-\sinh(at)\left[\cn(\omega t;k)S_2+\sn(\omega t;k)S_1\right].
\ee
Note that $S_3$ is the energy of the unperturbed harmonic oscillator.
Clearly, this quantity explodes for large times. Hence, the time-dependent
harmonic oscillator described by (\ref {more:ham}) is at resonance.

%%%%%%%%%%%%%%%%%%%%%%%%%%%%%%%%%%%%%%%%%%%%%%%%%%%%%%%%%%%%%%%%%%%%%%%%%%%%%%%%%
\section{Discussion}

In Part I of this work a method was developed to solve time-dependent
Hamiltonians with the assumption that they equal a time-dependent
linear combination of generators of a finite Lie algebra.
The method aims at finding all solutions, given one special solution.
This Part II demonstrates how the method of Part I can be
applied in two well-known cases, one corresponding with SU(2)
symmetry, the other with SU(1,1). In this way, a number of known results
are treated in a unified way. But in both cases the method
is shown to produce new results as well.

%other Lie algebras
We did not try to reproduce the most general results
found in the literature. We are confident that we could
do so, at the expense of writing a more technical and less
pedagogical paper. Of more interest is the application
of our method to other Lie algebras.

In \cite {LKJ96}, the Lie algebra that we used in Section 4 is
extended to contain 6 elements $\frac 12P^2$, $\frac 12Q^2$, $\frac 12(PQ+QP)$, $P$, $Q$, $\Io$.
With this extension it becomes possible to calculate the time evolution in the Heisenberg
picture of physically interesting quantities such as the position $Q$ and
the momentum $P$. One then can calculate the classical phase portrait
by studying the orbit $t\rightarrow \langle\psi|Q(t)|\psi\rangle,\langle\psi|P(t)|\psi\rangle$
for an arbitrary wavefunction $\psi$. This will be done in a future work.

Weigert \cite {WS97} has considered the general case of SU(N) symmetry.
Lopez and Suslov \cite {LL07} used the Heisenberg–Weyl group N(3) to describe a
forced harmonic oscillator.
Finally, Cari\~nena et al  \cite {CDR08}, among others, are interested
in developing a superposition principle for nonlinear equations by
mapping the solutions onto the solutions of linear equations with time-dependent
coefficients.

%\section*{}
%%%%%%%%%%%%%%%%%%%%%%%%%%%%%%%%%%%%%%%%%%%%%%%%%%%%%%%%%%%%%%%%%%%%%%%%%%%

\label{lastpage}

\end{document}